\documentclass[prl,twocolumn,aps,showpacs,floats,letterpaper,floatfix,groupedaddress,preprintnumbers]{revtex4}

\usepackage{graphicx,epsf, epsfig, amssymb}
\usepackage{bm}
\usepackage{longtable}

\def\be{\begin{equation}}
\def\ee{\end{equation}}
\def\beq{\begin{eqnarray}}
\def\eeq{\end{eqnarray}}

\begin{document}

\preprint{NSF-KITP-06-08}

\title{Rayleigh-Plateau and Gregory-Laflamme instabilities of black strings}

\author{Vitor Cardoso} \email{vcardoso@phy.olemiss.edu}
\affiliation{Department of Physics and Astronomy, The University of Mississippi, University, MS 38677-1848, USA
\footnote{Also at CFC
Universidade de Coimbra,
Portugal}}

\author{\'Oscar J. C. Dias\footnote{Also at KITP
UCSB, Santa Barbara, CA 93106, USA.}} \email{odias@perimeterinstitute.ca}
\affiliation{Perimeter Institute for Theoretical Physics, Waterloo, Ontario N2L 2Y5, Canada  \\and
Department of Physics, University of Waterloo, Waterloo, Ontario N2L 3G1, Canada }

\date{\today}

\begin{abstract}
Many and very general arguments indicate that the event horizon
behaves as a stretched membrane. We explore this analogy by
associating the Gregory-Laflamme instability of black strings with a
classical membrane instability known as Rayleigh-Plateau instability.
We show that the key features of the black string instability can be
reproduced using this viewpoint. In particular, we get good agreement
for the threshold mode  in all dimensions and exact agreement for large
spacetime dimensionality. The instability timescale is also well described
within this model, as well as the dimensionality dependence. It also predicts
that general non-axisymmetric perturbations are stable.
We further argue that the instability of ultra-spinning black holes
follows from this model.
\end{abstract}

\maketitle
The existence of black holes is perhaps the most dramatic prediction of Einstein's theory, the very concept of
which makes full use of the non-linearity of the equations and also of our notion of space and time. Despite
being apparently very complex objects, black holes can be associated with many of the familiar quantities of
everyday physics. The first major breakthrough in this direction was hinted at by Bekenstein \cite{bekenstein},
who conjectured that black holes are endowed with thermodynamic properties, namely with an entropy proportional
to its area. That black holes are thermodynamics entities was established once and for all by Hawking
\cite{hawking}, who verified explicitly that black holes radiate, and therefore have an associated temperature.
This analogy carries over to higher dimensional scenarios, which seem to be a prerequisite for consistency in
many modern theories. In the general higher dimensional gravity theories, there are other objects called black
branes \cite{horowitzstrominger}, with event horizons. These are basically extended black holes: the horizon,
instead of having the topology of a sphere, can have for instance the topology of sphere times a line -- a
cylinder. The analogy with thermodynamics can still be formulated, and in fact one can even go a further and
formulate an analogy with hydrodynamics \cite{kovtun1}. The thermodynamic description, both for black holes and
black branes, is based on the  four laws of black hole dynamics formulated by Bardeen, Carter and Hawking
\cite{BardCartHawk}. The first law (we will take for simplicity uncharged, static objects) describes how a
black hole, characterized by its mass $M$, horizon area $A$, and $T=1/(32 \pi M)$, evolves when we throw an
infinitesimal amount of matter into it:
\be dM=TdA \,.\label{firstlaw}\ee
The second law states that in any classical process the horizon
area must increase, $dA\geq 0$. It is very tempting to associate
these two laws with the first and second laws of thermodynamics,
respectively, in which case $T$ would be proportional to a
temperature and $A$ to an entropy (the other two laws of black
hole mechanics also have a correspondence with the zero and fourth
thermodynamic laws). The final ingredient to proceed consistently
with this association was given by Hawking \cite{hawking}, who
realized that black holes are indeed radiating objects and that
one can indeed associate them with a temperature $T_{\rm H}=4T$.

We can also argue that Equation (\ref{firstlaw}) can be looked at as a law for fluids, with $T$ being an
effective surface tension \cite{booktension}. Regarding the event horizon as a kind of fluid membrane is a
position adopted in the past \cite{membraneparadigm}. The first works in black hole mechanics actually
considered $T$ as a surface tension (see the work by Smarr \cite{smarr} and references therein), which is
rather intuitive: in fluids the potential energy, associated with the storage of energy at the surface, is
indeed proportional to the area. Later, Thorne and co-workers \cite{membraneparadigm} developed the ``membrane
paradigm'' for black holes, envisioning the event horizon as a kind of membrane with well-defined mechanical,
electrical and magnetic properties. Not only is this a simple picture of a black hole, it is also useful for
calculations and understanding what black holes are really like. There are other instances where a membrane
behavior seems evident: Eardley and Giddings \cite{eardley}, studying high-energy black hole collisions found a
soap bubble-like law for the process, while many modern interpretations for black hole entropy and gravity
``freeze'' the degrees of freedom in a lower dimensional space, in what is known as holography \cite{holo}. In
\cite{kovtun1} it was shown that a ``membrane'' approach works surprisingly well, yielding precisely the same
results as the AdS/CFT correspondence. There have also been attempts to work the other way round: computing
liquid surface tension from the (analog) black hole entropy \cite{surfacetbh}.

\begin{figure}[h]
\centerline{\includegraphics[width=7.5 cm,height=3 cm] {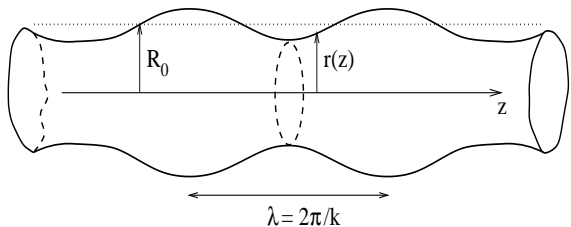}} \caption{Black strings and fluid cylinders
are unstable to perturbations on the extended dimension, i.e., along the axis of the cylinder. Ripples
propagating along this axis grow exponentially with time for wavelengths of order of the radius of the
cylinder.} \label{fig:inst}
\end{figure}

We will consider here an interesting class of black objects
without spherical topology, the black branes, and treat them as
fluids with a surface tension subjected to (\ref{firstlaw}) and
without gravity. A broad class of these objects are unstable
against gravity, in a mechanism known as Gregory-Laflamme
instability \cite{gl}. This is a gravitational instability along
the ``extended'' dimension, pictured in Figure \ref{fig:inst}. We
will show that we can mimic most aspects of this instability using
the fluid analogy.

Take a black string with radius $R_0$, and extended along a direction $z$. Then the Gregory-Laflamme mechanism
makes any small perturbation with wavelength $\lambda$ of the order of, or larger than, the radius of the
cylinder $R_0$ grow exponentially with time. This is a very robust long-wavelength instability.
The details for the instability can be found in the original papers \cite{gl}. For wavelengths larger than a
threshold $\lambda_c$, the instability appears. In the general setup of black strings with $D-1$ spatial
directions with radius $R_0$ and a transverse direction $z$, one finds \cite{gl} that the threshold wavenumber
increases with dimension number $D$ and so does the maximum growth timescale. For very large number of
dimensions, the threshold mode behaves as \cite{kol}
\be k_c \equiv 2\pi/\lambda_c \sim \sqrt{D}/R_0 \label{GLkol}\,,
\ee
%
Quite remarkably, we can find a similar instability for fluids
with surface tension, as shown in Plateau's celebrated study
\cite{plateau} on the stability of bodies under the influence of
surface tension. He established a fundamental result of classical
continuum mechanics: a cylinder longer than its circumference is
energetically unstable to breakup. This result was put on a firmer
basis by Lord Rayleigh \cite{rayleigh}, who computed the exact
instability timescale for the problem. The reasoning for the
appearance of an instability is the following: consider a small
disturbance of a long cylinder (we take its axis as the $z-$axis)
of fluid with radius $R_0$ and height $z$. Considering a small
axisymmetric perturbation along the surface of the cylinder, we
write for the disturbed cylinder
\be r(z)=R_0+\epsilon \,R_1\cos (kz)+\epsilon^2 R_2 \,,\label{rperturbation}\ee
where $\epsilon$ measures the perturbed quantities ($R_2$ is a second order quantity, and its usefulness will
be understood shortly). The volume of this cylinder can be easily computed to be
\be V= z \pi\left [R_0^2+\epsilon^2\left (2R_0R_2+\frac{R_1^2}{2} \right )\right ]\,+{\cal O}(\epsilon^3)\,.\ee
If we impose constant density one must have $R_2=-\frac{R_1^2}{4R_0}$. With this condition, the surface area of
the disturbed cylinder is
\be A=z\pi\left [ 2R_0+\frac{\epsilon^2R_1^2}{2R_0}\left (k^2R_0^2-1 \right )\right ] \,.\ee
The potential energy per unit length is therefore
\begin{eqnarray}
 P=\frac{\pi \epsilon^2 R_1^2}{2R_0}\left (k^2R_0^2-1\right )T \,.
 \label{potentialE}
\end{eqnarray}
We conclude that the system is unstable for $k<1/R_0$, since in this case a perturbation of the form
(\ref{rperturbation}) decreases the potential energy.

One can further show that non-axisymmetric perturbations, with
profile $r(z,\phi)=R_0+\epsilon \,R_1\cos (kz) \cos
(m\phi)+\epsilon^2 R_2$ (where $m$ is an integer that identifies
the angular mode) are stable for any $m\neq 0$ \cite{rayleigh}.
The reason is that the potential energy for these non-axisymmetric
modes is given by $P=\frac{\pi \epsilon^2 R_1^2}{2R_0}\left
(k^2R_0^2-1+m^2\right )T$, which never decreases for $m\neq 0$.

We can generalize the Rayleigh-Plateau construction to a general number of dimensions. Take a hyper-cylinder
with $D-1$ spatial directions with radius $R_0$ and a transverse direction $z$ (for the previous example
$D=3$). The axisymmetric threshold wavenumber is
\be R_0 k_c=\sqrt{D-2} \qquad {\rm with}\:\: k_c \equiv
2\pi/\lambda_c \,.\label{critk}\ee
For wavenumbers smaller (larger wavelengths) than this critical
value the cylinder is unstable. Moreover, as in the original
Rayleigh-Plateau situation, only symmetric modes seem to be
unstable; non-axisymmetric modes are in general stable. Therefore,
the Rayleigh-Plateau instability, like the Gregory-Laflamme
instability, should disappear for modes other than the $s-$modes.
This was recently conjectured  by Hovdebo and Myers
\cite{hovdebomyers} using an argument based on the relation
between the thermodynamic and the Gregory-Laflamme instabilities
\cite{reall}. Kudoh \cite{kudoh} has explicitly verified that the
Gregory-Laflamme instability only affects $s-$modes.

\begin{figure}[h]
\centerline{\includegraphics[width=6.5 cm,height=4 cm] {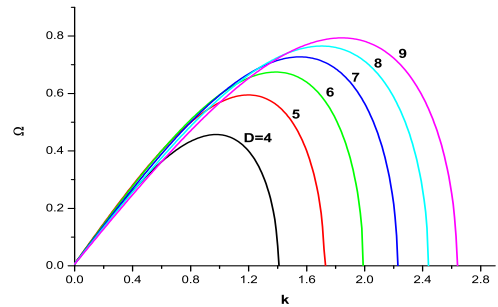}} \caption{Rayleigh-Plateau instability
of a hyper-cylinder in several dimensions. Here the effective surface tension and density were chosen to match
those of a higher-dimensional non-rotating black hole.
} \label{fig:insttime}
\end{figure}

To motivate quantitatively  the suggested association between the Rayleigh-Plateau and the Gregory-Laflamme
instabilities, it is important to compare the dependence of the threshold wavenumber $k R_0$ on dimension $D$,
for both instabilities. This is done by comparing (\ref{GLkol}) with (\ref{critk}). In Table
\ref{tab:thresholdmode}, we list the value of the Rayleigh-Plateau threshold mode $R_0 k_c$ for several
dimensions $D$. We also list the threshold wavenumber for the Gregory-Laflamme instability, with values taken
from \cite{kol}. There is good agreement between them. In the large $D$-limit there is exact agreement: both
the Gregory-Laflamme and the Rayleigh-Plateau critical wavenumber behave as $k_cR_0 \sim \sqrt{D}$. This is, we
think, a non-trivial check on the conjecture that black branes behave as fluid membranes with surface tension.
\begin{table*}
\caption{\label{tab:thresholdmode} Dimensionless threshold
wavenumber $k R_0$ for the Rayleigh-Plateau instability of a
higher dimensional fluid cylinder, and the corresponding threshold
wavenumber for the Gregory-Laflamme instability (data taken from
\cite{kol}).
 }
\begin{ruledtabular}
\begin{tabular}{ccccccccc}  \hline
$D \:\: {\rm spatial}\:\: {\rm dimensions}$ &4&5&6&7&8&9&49&99\\
\hline
{\rm Rayleigh-Plateau}&1.41     &1.73  &2.00 &2.24 &2.45&2.66&6.78&9.80\\ 
{\rm Gregory-Laflamme}&0.876 &1.27 &1.58  &1.85 &2.09&2.30&6.72&9.75\\ 
\end{tabular}
\end{ruledtabular}
\vskip -2mm
 \label{tabela}
\end{table*}
We can further compare the evolution of the instability timescale with its wavelength, and study the dependence
of the instability timescale on the spacetime dimension. We compute the Rayleigh-Plateau instability timescale
using fluid dynamics, following Rayleigh \cite{rayleigh}. For a 3-dimensional cylinder, and assuming the
perturbation goes as $R_1 \sim e^{\Omega t}$, Rayleigh gets the following expression for $\Omega$:
\be \Omega^2=\frac{T}{\rho R_0^3}\frac{i k R_0 J'_0(ikR_0)}{J_0(ikR_0)}
\left (1-k^2R_0^2\right )\,, \ee
with $J$ being a Bessel function. We have generalized Rayleigh's procedure for higher dimensions, and the
results are shown in Figure \ref{fig:insttime}. We assumed an effective surface tension and density associated
to the temperature and energy density of a $D$-dimensional Schwarzschild black hole. The qualitative behavior
of the instability timescale matches surprisingly well that of the Gregory-Laflamme \cite{gl} one. In
particular note that (i) the maximum growth rate grows with the number of spacetime dimensions, and (ii) the
corresponding wavenumber also grows with dimension number $D$. Moreover, the location of the threshold
wavenumber is exactly as predicted with the energy argument (see eq. (\ref{potentialE})). There is one
discrepancy only: the maximum instability for the Rayleigh-Plateau case, i.e.,  the maximum $\Omega$ is
approximately one order of magnitude larger than the maximum of the Gregory-Laflamme. This could be due to the
complete neglect of gravity in the outside of the cylinder (in the Rayleigh-Plateau analogy). Indeed if one
included gravity effects, a redshift was bound to occur, thereby lowering $\Omega$ \cite{rob}.

It is interesting to ask what is the endpoint of the Rayleigh-Plateau instability.
 In four-dimensions, the Rayleigh-Plateau
instability makes a cylinder pinch off, giving rise to drop
formation (see for instance the original works
\cite{plateau,rayleigh} or the review work by Eggers
\cite{eggers}). To see that this is expected, consider a $D=3$
cylinder breaking up at a Rayleigh-Plateau length, i.e., a
cylinder of height $z=\lambda_{\rm c}=2\pi R_0$. Can this give
rise to a drop of the same volume? Same volume implies that the
radius of the final spherical drop would be $1.6765R_0$. This
makes the surface area of the drop smaller than the corresponding
for the cylinder, $S_{\rm drop}=0.8947S_{\rm cyl}$, and thus it
makes sense that a drop is the final state. Now, if we play this
geometrical game in higher dimensions, we conclude that there is
critical dimension between $D=10$ and $D=11$. For $D\geq 11$ a
sphere seems not to be the favored endpoint, since it no longer
has less surface area. Once again, there is a very similar
phenomena in the gravity case (see ,e.g., the discussion in
\cite{hovdebomyers,Sorkin,park}). Take a black string breaking up
at a Gregory-Laflamme length. Sorkin found a critical spacetime
dimension ($d=D+1$) between $d=13$ and $d=14$, above which the
black string is no longer entropically unstable against the
formation of a spherical black hole \cite{Sorkin}.
 It would be
interesting to further study this issue. In particular, it is
important to understand what is the endpoint of the
Rayleigh-Plateau instability for $D\geq 11$, and to find if in the
fluid model there is a new branch of solutions that would be the
analogue of the non-uniform black string solutions of
\cite{nonUnif}, and if so to study their stability.

It has been shown by Horowitz and Maeda \cite{horomaeda} that pinch-off, if it occurs at all in the black
string case, must do so in infinite affine time. This immediately suggests that an attractive endpoint could be
non-uniform black strings \cite{horomaeda,nonUnif}. Now, the breakup of liquid jets with surface tension in the
absence of viscosity is known to happen in finite time, but the inclusion of viscosity (which seems a necessary
ingredient to model realistic black objects \cite{wilczek}) may change this \cite{eggers}, so even this
unexpected feature might be discussed within this analogue model. It is quite amusing that some of the
dynamical features of the instability of black strings have already been observed in liquids: the final state
of some liquid bridges (finite-size liquid cylinders), unstable under the Rayleigh-Plateau instability, is a
non-symmetric state (see \cite{liquidbridgesstate} and its references) and the breakup of liquid jets is quite
generally a self-similar phenomena \cite{eggers}.

The interpretation of $T$ as surface tension can also improve -- and strengthen -- our understanding of the
instability of ultra-spinning black holes (Myers-Perry \cite{myersperry} black holes with high rotation), a
conjecture recently made by Emparan and Myers in connection with the Gregory-Laflamme instability
\cite{myersemparan}. Take a slowly rotating black hole. As the rotation rate increases, the surface becomes
flatter at the poles until zero Gaussian curvature occurs. But a liquid drop (at least in four dimensions)
develops instabilities before zero Gaussian curvature is reached \cite{chandra}. Assuming a correspondence
between rotating liquid drops and rotating black holes, one may well expect ultra-spinning black holes to be
unstable. We note that this very same reasoning was applied by Smarr \cite{smarr} many years ago. In four
dimensions, it seems that the upper Kerr bound in the angular momentum is small enough to avoid the development
of such instabilities. However, in dimensions higher than six, rotating black holes have no Kerr-like bound,
and the instability might well set in.

It is possible that an event horizon behaves dynamically much as a fluid interface without gravity, as we have
shown. If this is so, the second law of black hole dynamics should be something like a soap-bubble law,
sphericity is preferred for it is the minimum energy shape (which would justify why most stable solutions in GR
have spherical topology). Finally, there are many ways to extend these results, by including additional
effects, like charge or rotation in the problem. One would naively expect either of these to increase the
instability, but a more careful study has to be done. Take rotation for instance. The centrifugal force,
scaling with $r$, contributes with a de-stabilizing effect, since there is an increased pressure under a crest
but a reduced pressure under a trough. So in principle, the threshold wavelength should be smaller
\cite{interfacial}. However, there is no simple reasoning (as far as we know) to tell the effects of rotation
on the strength of the perturbation, i.e., on the instability timescale. This is partly because the effective
surface tension of black hole itself depends on rotation, decreasing with increasing rotation. This would be an
interesting direction for further study.

We would like to thank the participants of the KITP Program: ``Scanning New Horizons: GR Beyond 4 Dimensions",
in particular E. Berti, L. Bombelli, M. Cavagli\`a, J. Hovdebo, D. Marolf, R. Myers, E. Sorkin and T. Wiseman
for valuable comments and suggestions.
VC acknowledges financial support from Funda\c c\~ao Calouste
Gulbenkian, and OD from FCT (grant SFRH/BPD/2004), and from a
NSERC Discovery grant (University of Waterloo).
Research at Perimeter Institute is supported in part by funds from
NSERC of Canada and MEDT of Ontario.
This research was supported by the NSF under Grant No.
PHY99-07949.


\begin{thebibliography}{99}

\bibitem{bekenstein} J. D. Bekenstein, Phys. Rev. D {\bf 7}, 2333 (1973).

\bibitem{hawking} S. W. Hawking, Commun. Math. Phys. {\bf 43}, 199 (1975).

\bibitem{horowitzstrominger} G. T. Horowitz and A. Strominger, Nucl. Phys. B {\bf 360},
197 (1991).

\bibitem{kovtun1} P. Kovtun, D. T. Son and A. O. Starinets, Phys. Rev. Lett. {\bf 94}, 111601 (2005);
JHEP {\bf 0310}, 064 (2003).

\bibitem{BardCartHawk} J.~M.~Bardeen, B.~Carter and S.~W.~Hawking,
Commun. Math. Phys. {\bf 31}, 161 (1973).

\bibitem{booktension} R. Defay, I. Prigogine, A. Bellemans and D. H. Everett, {\it Surface tension and
adsorption}, (Joun Wiley and Sons, New York, 1966).

\bibitem{membraneparadigm} K. S. Thorne, D. A. MacDonald and R. H. Price,
{\it Black Holes: The Membrane Paradigm} (Yale University Press, 1986).

\bibitem{smarr} L. Smarr, Phys. Rev. Lett. {\bf 30}, 71 (1973).

\bibitem{eardley} D. M. Eardley and S. B. Giddings, Phys. Rev. D {\bf 66}, 044011 (2002).

\bibitem{holo} G. 't Hooft, gr-qc/9310026; L. Susskind, J. Math. Phys. {\bf 36}, 6377 (1995);
J. M. Maldacena, Adv. Theor. Math. Phys. {\bf 2}, 231 (1998); E. Witten, Adv. Theor. Math. Phys. {\bf 2}, 253
(1998); R. Bousso, JHEP {\bf 9907}, 004 (1999).

\bibitem{surfacetbh} D. J. E. Callaway, Phys. Rev. E {\bf 53}, 3738 (1996).

\bibitem{gl} R. Gregory and R. Laflamme, Phys. Rev. Lett. {\bf 70}, 2837 (1993); Nucl. Phys. B {\bf 428}, 399 (1994).

\bibitem{kol} B. Kol and E. Sorkin, Class. Quant. Grav. {\bf 21}, 4793 (2004).

\bibitem{plateau} J. Plateau, {\it Statique Exp\'erimentale et Th\'eorique des Liquides
Soumis aux Seules Forces Mol\'eculaires}, (Paris, Gauthier-Villars, 1873).

\bibitem{rayleigh} L. Rayleigh, Proc. Lond. Math. Soc. {\bf 10}, 4 (1878).

\bibitem{hovdebomyers} J. L. Hovdebo and R. C. Myers, hep-th/0601079.

\bibitem{reall} H. S. Reall, Phys. Rev. D {\bf 64}, 044005 (2001).

\bibitem{kudoh} H. Kudoh, hep-th/0602001.

\bibitem{rob} R. C. Myers, private communication.

\bibitem{eggers} J. Eggers, Rev. Mod. Phys. {\bf 69}, 865 (1997).

\bibitem{Sorkin} E.~Sorkin, Phys. Rev. Lett.  {\bf 93}, 031601 (2004).

\bibitem{park} M-I. Park, Class. Quant. Grav. {\bf 22}, 2607 (2005).

\bibitem{nonUnif} S.~S.~Gubser, Class. Quant. Grav.  {\bf 19}, 4825 (2002);
T.~Wiseman, Class.\ Quant.\ Grav.\  {\bf 20}, 1137 (2003).

\bibitem{horomaeda} G. T. Horowitz and K. Maeda, Phys. Rev. Lett. {\bf 87}, 131301 (2001).

\bibitem{wilczek} M. Parikh and F. Wilczek, Phys. Rev. D {\bf 58}, 064011 (1998).

\bibitem{liquidbridgesstate}
J.M. Montanero, G. Cabezas and J. Acero, Phys. Fluids {\bf 14},
682 (2002).

\bibitem{myersperry} R. C.~Myers and M. J.~Perry,
Ann. Phys.\ {\bf 172}, 304 (1986).

\bibitem{myersemparan} R. Emparan and R. C. Myers,
JHEP {\bf 0309}, 025 (2003).

\bibitem{chandra} S. Chandrasekhar, Proc. Roy. Soc., Ser. A {\bf 286}, 1 (1964).

\bibitem{interfacial} L. E. Johns and R. Narayanan, {\it Interfacial instability},
(Springer-Verlag, New York, 2002).

\end{thebibliography}
\end{document}